\documentclass[lettersize,journal]{IEEEtran}
\usepackage{amsmath,amsfonts}
\usepackage{algorithmic}
\usepackage{algorithm}
\usepackage{array}
\usepackage[caption=false,font=normalsize,labelfont=sf,textfont=sf]{subfig}
\usepackage{textcomp}
\usepackage{stfloats}
\usepackage{url}
\usepackage{svg}
\usepackage{verbatim}
\usepackage{graphicx}
\usepackage{cite}
\usepackage{tikz}            
\usepackage{pgfplots}        
\pgfplotsset{compat=1.18}    
\usepgfplotslibrary{fillbetween}
\usepackage{pgfplotstable}
\usetikzlibrary{patterns}

\usepackage{pifont}

\usepackage{booktabs}
\usepackage{makecell}
\usepackage{multirow}
\usepackage{multicol}
\usepackage{adjustbox}
\usepackage[table]{xcolor}
\usepackage{threeparttable}

\definecolor{phoneacc}{RGB}{0,70,140}
\definecolor{darkgreen}{RGB}{0,120,0}
\newcommand{\pacc}[1]{\textcolor{phoneacc}{#1}}
\newcommand{\noaudio}{\textcolor{gray}{ -- }}

\usepackage{orcidlink}
\hypersetup{
    hidelinks,        
    colorlinks=false, 
}

\begin{document}

\title{Cross-Modal Masking for Robust Silent\\Speech Synthesis Using sEMG and Lipreading}

\author{Eder del Blanco\textsuperscript{\orcidlink{0000-0001-6510-778X}, $\dagger$}, David Gimeno-Gómez\textsuperscript{\orcidlink{0000-0002-7375-9515}, $\dagger$}, Eva Navas\textsuperscript{\orcidlink{0000-0003-3804-4984}}, Carlos-D. Martínez-Hinarejos\textsuperscript{\orcidlink{0000-0002-6139-2891}}, Inma Hernáez\textsuperscript{\orcidlink{0000-0003-4447-7575}}
\thanks{E. del Blanco, E. Navas, and I. Hernáez are with the Aholab research group within the HiTZ Center at University of the Basque Country (UPV/EHU), Barrio Sarriena s/n, 48940 Leioa, Bizkaia, Spain (e-mail: eder.delblanco@ehu.eus; eva.navas@ehu.eus; inma.hernaez@ehu.eus). D. Gimeno-Gómez and C.-D. Martínez-Hinarejos are with the PRHLT research center, Universitat Politècnica de València (UPV), Camino de Vera s/n, 46022, València, Spain (e-mail: dagigo1@dsic.upv.es; cmartine@dsic.upv.es).}
\thanks{\textsuperscript{$\dagger$}These authors contributed equally to this work.}%
\thanks{Manuscript received Month Day, Year; revised Month Day, Year.}}

\markboth{Journal of Transactions on Audio, Speech and Language Processing,~Vol.~XX, No.~X, Month~Year}%
{del Blanco \MakeLowercase{\textit{et al.}}: Cross-Modal Masking for Robust Silent Speech Synthesis Using sEMG and Lipreading}

\IEEEpubid{0000--0000/00\$00.00~\copyright~2026 IEEE}

\maketitle

\begin{abstract}
Speech restoration through silent speech interfaces (SSIs) has emerged as a promising assistive technology for individuals with impaired or absent laryngeal voice production. Among non-invasive SSI modalities, surface electromyography (sEMG) and video-based lipreading provide complementary articulatory information, yet their integration for continuous speech synthesis remains underexplored. Moreover, existing multimodal approaches rarely address robustness to modality degradation or temporary sensor failure, limiting their applicability in realistic scenarios. In this work, we propose a masked multimodal speech synthesis framework that jointly leverages sEMG and lipreading signals through modality masking during training. Under multi-speaker settings, the proposed approach reduces word error rate by up to 14 absolute percentage points compared to the strongest unimodal baseline. Experimental results not only show that masking strategies are critical for these performance gains and robustness under low-bitrate conditions, but also that they generalize better than degradation-specific data augmentations in the presence of modality absence conditions. Phone-level analyses further reveal complementary contributions across modalities, with particularly strong benefits for vowels and for specific consonant groups. Overall, these findings demonstrate the effectiveness and robustness of masked multimodal integration for silent speech synthesis, although adaptation to laryngectomized speakers remains an open research challenge.
\end{abstract}

\begin{IEEEkeywords}
Silent speech interfaces, speech synthesis, surface electromyography, lipreading, masked multimodal learning.
\end{IEEEkeywords}

\section{Introduction}
\label{sec:intro}
\IEEEPARstart{S}{peech} loss following total laryngectomy has a profound impact on the lives of affected individuals, often leading to feelings of social withdrawal and isolation~\cite{danker2010social}, which may ultimately contribute to the onset of clinical depression~\cite{byrne1993depression}. These effects are also closely tied to identity loss due to the inability to use one’s own voice~\cite{shadden2005identity} --- a direct consequence of vocal fold removal, which eliminates the primary source of voicing, pitch control, and loudness modulation.

To restore communication, several alaryngeal speech rehabilitation methods have been developed, including esophageal speech, electrolarynx-based speech, and tracheoesophageal voice prostheses. However, these approaches not only yield reduced naturalness and intelligibility compared to laryngeal speech, but also may require medical intervention and extensive user training, limiting their effectiveness in everyday communication~\cite{most2000acoustic,tang2015voice}. These limitations have motivated research into silent speech interfaces (SSIs), an assistive technology that aims to reconstruct speech from non-acoustic biosignals generated during speech production~\cite{gonzalez2020review}.

\begin{figure}[t]
  \centering
  \includegraphics[width=1.0\columnwidth]{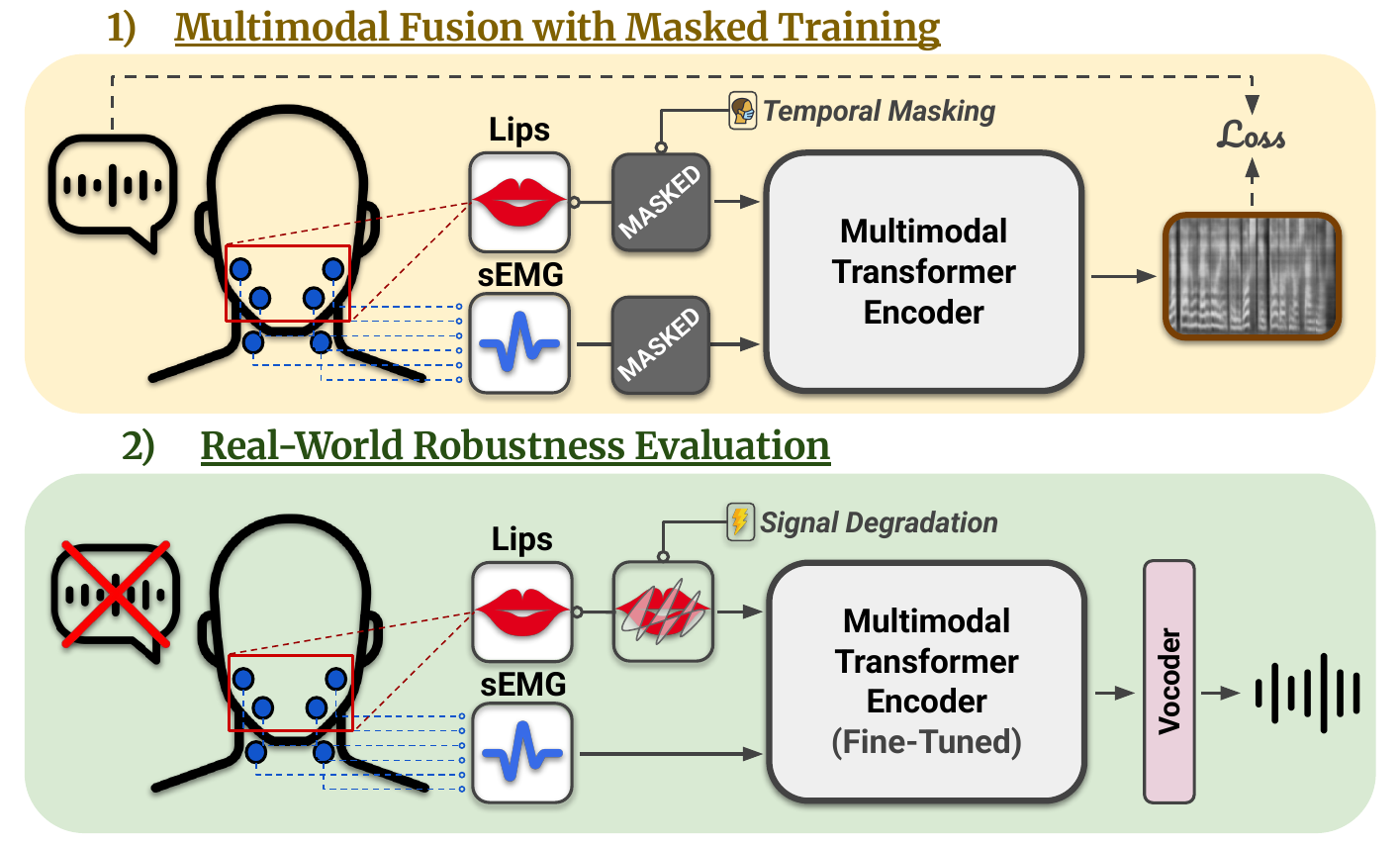}
  \caption{\textbf{Illustrative overview of the proposed multimodal speech synthesis framework.} Top: Masked multimodal fusion training enables the model to integrate lipreading and sEMG cues effectively. Bottom: Once trained, the model is fine-tuned and evaluated under simulated signal degradations (e.g., reduced video frame rate), demonstrating that the multimodal model maintains stronger robustness in real-world use cases.}
  \label{fig:overview}
\end{figure}

\IEEEpubidadjcol
Existing SSI approaches rely on diverse sensing modalities, many of which require specialized or intrusive equipment, such as ultrasound imaging~\cite{gosztolya2025silent} or electromagnetic articulography with intraoral sensors~\cite{teplansky2025articulatory}. In contrast, surface electromyography (sEMG)~\cite{meltzner2017silent,gaddy2022voicing} and video-based lipreading~\cite{sen2021personalized,laux2023care} have gained increasing attention due to their non-invasive nature and potential for real-world deployment. While both modalities capture articulatory cues during speech production, they differ fundamentally in the type of information they observe. sEMG measures the electrical activity generated by facial muscles using electrodes placed on the skin~\cite{gaddy2020digital}, providing a direct but spatially coarse representation of muscular activation. Lipreading, on the other hand, infers speech content from visual cues such as lip and mouth movements extracted from video~\cite{ma2022visual}, offering rich detail of external articulators while remaining indirectly related to underlying motor activity.

Despite their promise, both modalities present inherent limitations. sEMG signals are sensitive to electrode placement, inter- and intra-speaker physiological variability, and potential signal drift over time~\cite{salomons2025electrode}. Visual speech signals, while benefiting from spatial normalization through face detection and alignment techniques, remain sensitive to environmental conditions such as lighting variations, head pose changes, and occlusions, which can obscure or distort key articulatory cues~\cite{hong2023watch}. Furthermore, both modalities have been shown to exhibit systematic differences between audible and silent speech production~\cite{gaddy2020digital,petridis2018visual}, introducing an additional source of variability that can further hinder model generalization~\cite{gonzalez2020review}.

Importantly, the origin and nature of these limitations are different and can therefore be considered complementary. This observation suggests that combining sEMG and visual information may provide a more robust representation of speech production, potentially compensating the weaknesses of each individual modality. This is consistent with prior findings in multimodal SSI, where combining heterogeneous sensing modalities, such as lipreading and ultrasound tongue imaging, has been shown to improve speech decoding performance~\cite{hueber2010development}. 

However, despite this potential, the integration of sEMG and lipreading cues remains largely underexplored, with existing work typically limited to sentence-level classification, without addressing speech synthesis, and often operating in audible (vocalized) speech conditions~\cite{zhou2025avespeech}. Within this line of work, robustness to realistic conditions where one modality may be degraded or unavailable due to sensor noise, misplacement, or low-bitrate transmission is also rarely addressed, motivating the development of multimodal approaches capable of handling missing or unreliable inputs.

In this sense, masking strategies have been widely used to improve robustness in the field of speech technologies, including audio speech recognition~\cite{park2020specaugment}, automatic lipreading~\cite{ma2022visual}, and more recently audio-visual speech recognition~\cite{shi2022learning,gimeno2025tailored}, where modality dropout and cross-modal masking encourages the learning of complementary multimodal representations.

Motivated by all these observations, we propose a masked multimodal SSI that integrates sEMG and lipreading for speech synthesis. As illustrated in Fig.~\ref{fig:overview}, our framework is designed to learn robust cross-modal representations under modality degradation and absence conditions.

Our main contributions are threefold:

\begin{itemize}

    \item \textbf{Complementary Silent Speech Cues.} We first support that sEMG and lipreading signals are complementary, and that their fusion substantially improves performance, reaching around 76\% phone accuracy and 40\% word error rate in challenging multi-speaker settings, with clear gains over the best-performing unimodal baseline.

    \item \textbf{Critical Role of Masking.} We demonstrate the impact of temporal adaptive masking in learning robust multimodal representations, showing that its benefits extend beyond missing modalities to signal degradations such as video frame downsampling, advancing toward real-world deployment of this technology in practical scenarios.

    \item \textbf{Phone-Level Multimodal Analysis.} We conduct a detailed phone-level analysis to investigate the contributions of multimodal fusion, revealing consistent but non-uniform gains across most articulatory classes. In particular, we show that incorporating sEMG provides complementary information for vowels and affricates, while remaining limited for more challenging distinctions among plosives and nasals.
    
\end{itemize}

\section{Related Work}
\label{sec:related}

This section overviews key studies related to the present work, covering speech synthesis from sEMG and lipreading signals, the research gap in multimodal SSIs, and the impact of masking strategies in the speech technologies field.

\vspace{0.15cm}
\noindent\textbf{sEMG Speech Synthesis.} While traditionally focused on automatic speech recognition \cite{denby2010silent,gonzalez2020review}, sEMG-based speech interfaces have demonstrated the feasibility of direct, open-vocabulary speech synthesis from articulatory muscle activity \cite{zahner2014conversion,janke2017emg, diener2018comparison}, thereby underscoring the potential of this modality for speech restoration. Benchmark datasets such as the one introduced by Gaddy and Klein~\cite{gaddy2020digital} have enabled significant progress, although their impact remains largely limited to single-speaker settings. In response to this limitation, recent efforts have explored data augmentation and self-supervised learning \cite{chen2025confidence}, as well as improved generative architectures such as diffusion-based models \cite{ren2024diff}. In addition, Soft Speech Units have been proposed to decouple linguistic content from speaker identity, enabling more flexible synthesis \cite{scheck2023multi}. Complementarily, multi-speaker datasets in English \cite{scheck2024cross} and Mandarin \cite{li2023silent} have further extended training beyond single-speaker settings, yet none include recordings from laryngectomized users. Despite advances, however, performance remains strongly affected by participant- and session-dependent variability, which limits generalization across speakers and recording conditions. This dependence is compounded by practical challenges such as electrode placement variability~\cite{salomons2025electrode}, which hinders intelligibility in realistic settings.

\vspace{0.15cm}
\noindent\textbf{Lip-to-Speech Approaches.} Lipreading has emerged as a promising non-invasive alternative for speech synthesis, offering a sensing modality that is less affected by inter-session variability, as exemplified in applications involving tracheotomized individuals~\cite{laux2023care}. It has been the advent of deep learning that has led to significant advances in lip-to-speech synthesis. While early approaches relied on autoencoders coupled with recurrent neural networks~\cite{akbari2018lipaudspec}, subsequent sequence-to-sequence frameworks improved performance in more realistic scenarios, including larger vocabularies and increased robustness to head pose variations~\cite{prajwal2020lipwav}. Mira et al.~\cite{mira2022svts} emphasized the limitations of increasingly complex architectures and proposed a more streamlined alternative based on Conformer encoders~\cite{gulati20conformer} with a linear projection head, demonstrating the benefits of combining similarity-based losses with spectral reconstruction objectives. In parallel, auxiliary text transcription tasks have been shown to further improve speech intelligibility~\cite{kim2023lip}. More recent developments are also exploring diffusion-based generative models~\cite{richter2025lipdiffuser}. However, lipreading is not exempt from limitations. One fundamental limitation is its sensitivity to visual conditions (such as lighting variations and occlusions). Additionally, similarly to sEMG, presents a noticeable dependence on speaker-specific articulatory patterns~\cite{cox2008challenge,gimeno2023comparing}. Albeit recent advances have partially addressed these issues~\cite{shi2022learning,gimeno2025tailored}, robustness in real-world conditions remains a challenge due to the fact that only part of the articulatory information is visible~\cite{duchnowski2000development}.

\vspace{0.15cm}
\noindent\textbf{Multimodality in Silent Speech Interfaces.}
Multimodality has been extensively studied due to its ability to provide robustness against environmental noise and integrate complementary information from diverse data sources~\cite{ngiam2011multimodal}. However, the specific combination of sEMG and lipreading remains largely underexplored in the context of SSIs, despite the complementary potential of both modalities. Early attempts have been explored in prior work~\cite{freitas2014enhancing,freitas2014multimodal}, but these approaches were limited to word-level recognition tasks and did not exploit the representation learning capabilities of modern deep learning architectures. Another notable exception is the work conducted by Zhou et al.~\cite{zhou2025avespeech}, where audio, sEMG, and lipreading signals were jointly integrated within a unified framework. While their approach improved robustness under noisy conditions, further highlighting the potential benefits of multimodal SSIs in realistic scenarios, it addressed speech-to-text generation and formulated the task as sentence-level classification over a fixed set of utterances rather than continuous speech generation, which considerably limits both its practical applicability and the investigation of complementary interactions between sEMG and lipreading.

Other interesting multimodal SSI approaches explored in the field include the following: integration of lipreading with ultrasound tongue imaging for speech synthesis, achieving phone recognition rates of around 60\% in a laryngectomized speaker using conventional speech decoding paradigms~\cite{hueber2010development}; introducing a dataset combining radar-frequency sensing with audio and video-based lipreading~\cite{ge2023comprehensive}; fusing muscular sEMG activity with neural EEG signals, improving intelligibility while offering insights into underlying speech mechanisms~\cite{li2023hybrid}. All these studies support the idea that multimodal integration provides complementary information across sensing modalities, motivating continued research into robust multimodal SSIs.

\vspace{0.15cm}
\noindent\textbf{Masking in Multimodal Speech Technologies.} Masking has been largely adopted as an effective strategy to improve temporal modeling and robustness in speech processing. One of the most representative approaches in the acoustic domain is SpecAugment~\cite{park2020specaugment}, which applied time and frequency masking directly to speech spectrograms. Subsequently, Ma et al.~\cite{ma2022visual} extended this idea to automatic lipreading by proposing an adaptive temporal masking strategy that replaces contiguous video frames with their average values within 1-second segments, arguing that such proportional masking improves the disambiguation of visually similar phones (visemes)~\cite{fernandez2017optimizing}.

The importance of masking becomes even more pronounced in multimodal settings, especially when one modality is substantially more informative and may dominate the learning process. This phenomenon is well known in audio-visual speech recognition (AVSR), where the audio stream can overshadow visual modality~\cite{afouras2018deep}. To mitigate such imbalance, several approaches have introduced not only severe degradations to the audio input~\cite{ma2023auto,gimeno2025tailored}, but also modality dropout strategies~\cite{shi2022learning}, in which one modality is entirely removed during training. Both these techniques have shown to prevent the model from ignoring the less informative input stream and encourage more balanced cross-modal learning.

\section{Methods}
\label{sec:methods}

\begin{figure*}[!t]
  \centering
  \includegraphics[width=\textwidth]{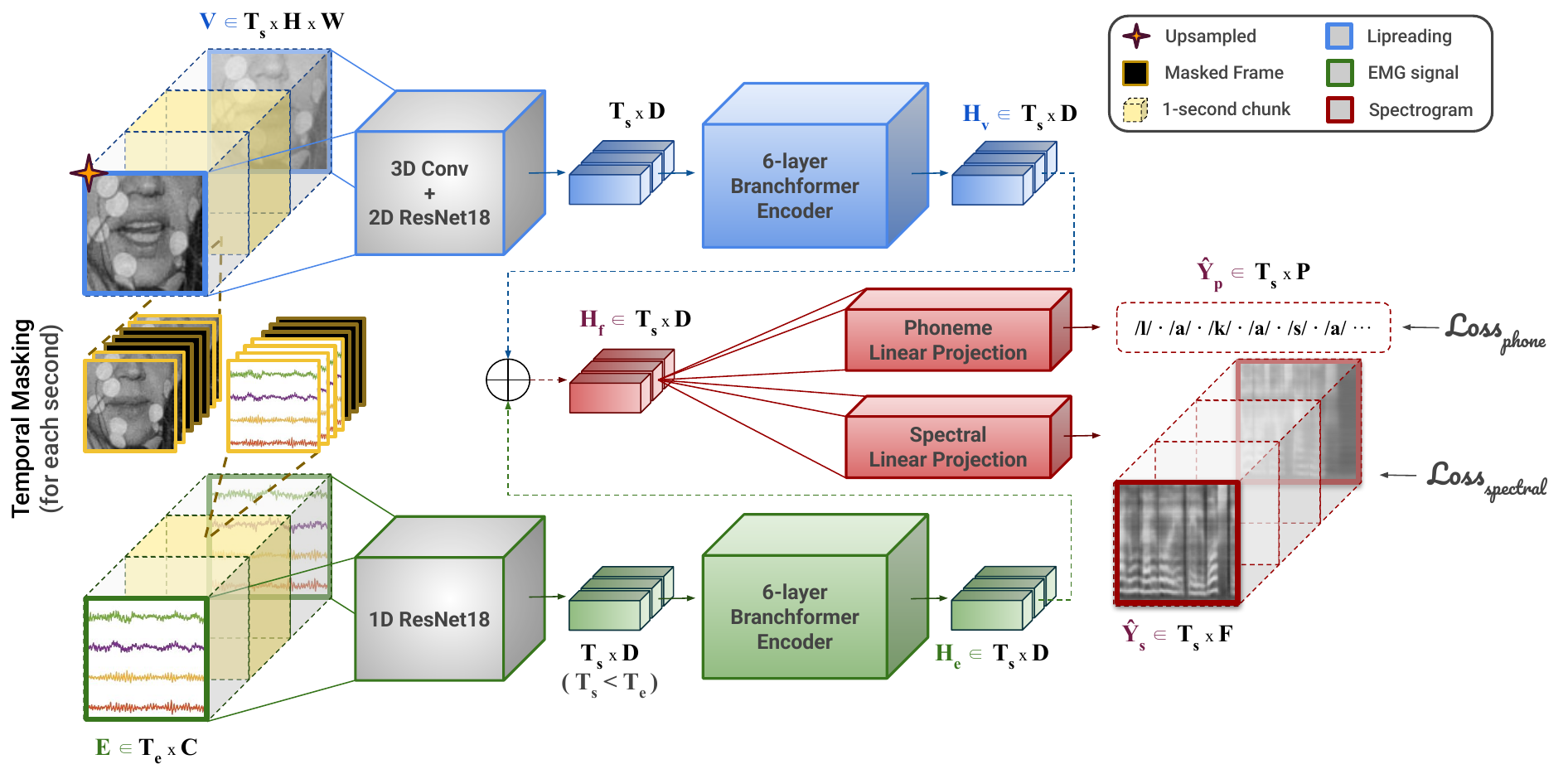}
  \caption{\textbf{Overall schema of the masked multimodal speech synthesizer.} The proposed framework jointly processes lip video frames (blue squares) and sEMG signals (green squares) using modality-specific encoders. For each 1-second chunk, temporal adaptive masking (orange squares) is applied independently to both modalities before feature extraction through convolutional-based frontends. The resulting embeddings are further encoded by separate 6-layer Branchformers to capture richer contextual dependencies, and subsequently fused via element-wise addition. The model is then optimized in a multi-task manner by feeding the multimodal representations into two task-specific projection heads for phone prediction (top red brach) and spectrogram estimation (bottom red branch).}
  \label{fig:method}
\end{figure*}

This section introduces the proposed multimodal speech synthesis framework, covering the problem formulation, the dual-stream architecture description, and the multimodal time masking strategy employed during training. An overview of the complete method is illustrated in Fig.~\ref{fig:method}.

\subsection{Problem Formulation}

Let $\mathcal{D} = \{(\mathbf{E}_i, \mathbf{V}_i, \mathbf{Y}_i)\}_{i=1}^{N}$ denote a multimodal dataset consisting of synchronized sEMG signals, lip video crops, and corresponding speech targets. For clarity, $(\mathbf{E}, \mathbf{V}, \mathbf{Y})$ denotes an arbitrary sample (or mini-batch) drawn from $\mathcal{D}$.

Each sEMG sequence is then represented as $\mathbf{E} \in \mathbb{R}^{B \times T_e \times C}$, corresponding to $C$-channel recordings sampled at high temporal  resolution. The lipreading modality is represented as $\mathbf{V} \in \mathbb{R}^{B \times T_v \times H \times W}$, where $H \times W$ denotes the spatial resolution of grayscale crops. In both cases, $B$ refers to the batch size, while $T_e$ and $T_v$ denote the raw number of temporal frames for sEMG and video, respectively.

Given this multimodal input, our objective is to reconstruct intelligible speech through a joint prediction of spectral and phonetic targets. We therefore define the speech supervision as $\mathbf{Y} = (\mathbf{Y}_s, \mathbf{Y}_p)$, where the spectral target is given by $\mathbf{Y}_s \in \mathbb{R}^{B \times T_s \times F}$, with $T_s$ denoting the number of temporal frames in the spectral representation and $F$ the number of frequency bins, and the phonetic alignment target by $\mathbf{Y}_p \in \mathbb{R}^{B \times T_s \times P}$, where $P$ denotes the size of the phone vocabulary.

To this end, beyond jointly optimizing spectral reconstruction and phonetic discrimination in a multi-task setting, we employ a dual-encoder attention-based architecture combined with structured cross-modal masking, collectively encouraging the model to learn complementary representations and maintain robust performance under realistic, noisy conditions.

\subsection{Model Architecture}
\label{subsec:model}

\vspace{0.15cm}
\noindent\textbf{sEMG Encoding Process.} Within the field of speech technologies, recent efforts have advanced end-to-end processing of raw signals~\cite{ma2021endtoend} and multimodal fusion~\cite{gimeno2025tailored}. Based on these insights, we define the sEMG encoder output sequence as:

\begin{equation}
\label{eq:emg_encoder}
\begin{split}
\mathbf{H}_e = f_e(\mathbf{E}), \qquad\qquad\\
\mathbf{H}_e = [\mathbf{h}_e^1, \dots, \mathbf{h}_e^{T'_e}] 
\in \mathbb{R}^{B \times T'_e \times D},
\end{split}
\end{equation}

\noindent where $f_e(\cdot)$ denotes the composition of a 1D ResNet-18 frontend~\cite{he2016deep}, which takes the raw sEMG input $\mathbf{E}$, followed by a Branchformer encoder~\cite{peng2022branchformer}. The ResNet frontend extracts local temporal patterns from the multi-channel sEMG signals while reducing the temporal resolution to exactly align with the target spectral frame rate $T_s$. To early capture longer-range muscle activation patterns, we modified the first convolutional layer to use a kernel size of 7, increasing the initial receptive field. Relative positional embeddings~\cite{dai2019transformerxl} are injected into the extracted $D$-dimensional frontend features to explicitly encode temporal order information before contextual modeling through a 6-layer Branchformer encoder. This Transformer-based variant was designed to jointly model local and global temporal dependencies through parallel self-attention\cite{vaswani2017attention} modules and gated convolutional blocks~\cite{sakuma2022mlpbased} --- a dual modeling particularly suitable for the nature of our speech cues. Preserving the feature dimensionality, it produces the contextualized hidden sEMG representation $\mathbf{H}_e$.

\vspace{0.15cm}
\noindent\textbf{Lipreading Encoding Process.} Similarly, the lipreading input $\mathbf{V}$ is first processed by a modified 2D ResNet-18 frontend. To better capture spatio-temporal features, we replaced the first convolutional layer with a 3D convolution spanning a 5-frame receptive field, consistent with previous work~\cite{ma2022visual}. After flattening the visual features along the temporal dimension, the 2D ResNet-18 extracts local visual patterns, producing $D$-dimensional embeddings. Relative positional embeddings are then added prior to a Branchformer encoder configured identically to its sEMG-stream counterpart. Together, these modules compose the lipreading processing, $f_v$, which outputs the contextualized visual representation denoted as:

\begin{equation}
\mathbf{H}_v = f_v(\mathbf{V}) = [\mathbf{h}_v^1, \dots, \mathbf{h}_v^{T'_v}] \in \mathbb{R}^{B \times T'_v \times D},
\end{equation}

\noindent where $T'_v$, unlike $T'_e$ for sEMG, is not obtained via convolutional striding but through a prior linear upsampling of the original video sequence (typically sampled at lower frame rates) to match the target spectral frame rate $T_s$.

\vspace{0.15cm}
\noindent\textbf{Multimodal Fusion.} This architecture design ensured temporal alignment between the modalities for the subsequent multimodal fusion, such that $T'_e = T'_v = T_s$. The aligned sEMG and lipreading hidden representations are then combined via an element-wise additive fusion strategy:

\begin{equation}
\mathbf{H}_f = \text{LayerNorm}(\mathbf{H}_e + \mathbf{H}_v),
\end{equation}

\noindent where $\mathbf{H}_f \in \mathbb{R}^{B \times T_s \times D}$ denotes the fused multimodal representation, with layer normalization applied to stabilize training. The rationale behind choosing an additive scheme was to produce a complementary hidden representation that remains robust even if one of the modalities is missing.

\vspace{0.15cm}
\noindent\textbf{Speech Synthesis Decoding.}
Inspired by recent multi-task schemas~\cite{kim2023lip}, the multimodal representation is 
then decoded into the target speech outputs, $\mathbf{Y}$, via two parallel linear projection heads.
The first head predicts spectral features:

\begin{equation}
\hat{\mathbf{Y}}_s = \mathbf{H}_f \mathbf{W}_s + \mathbf{b}_s \in \mathbb{R}^{B \times T_s \times F},
\end{equation}

\noindent where $\mathbf{W}_s \in \mathbb{R}^{D \times F}$ and $\mathbf{b}_s \in \mathbb{R}^{F}$ denote the learnable projection matrix and bias term for spectral reconstruction, and $F$ is the number of frequency bins.

The second head predicts frame-level phonetic labels:

\begin{equation}
\hat{\mathbf{Y}}_p = \mathbf{H}_f \mathbf{W}_p + \mathbf{b}_p \in \mathbb{R}^{B \times T_s \times P},
\end{equation}

\noindent where $\mathbf{W}_p \in \mathbb{R}^{D \times P}$ and $\mathbf{b}_p \in \mathbb{R}^{P}$ are the corresponding projection parameters, and $P$ denotes the size of the phone vocabulary. By jointly optimizing spectral reconstruction and phonetic discrimination within a unified decoding framework, we offer a more informed and acoustically aligned auxiliary supervision compared to transcription-based objectives.

\vspace{0.15cm}
\noindent\textbf{Loss Function.} Given this multi-task formulation, the proposed model is optimized using a composite objective:

\begin{equation}
\begin{aligned}
\mathcal{L}_{\text{total}} =& \mathcal{L}_{\text{spectral}}  + \lambda \mathcal{L}_{\text{phone}}, \\
\mathcal{L}_{\text{spectral}} &= \mathcal{L}_{mse} + \mathcal{L}_{\text{conv}},
\end{aligned}
\end{equation}

\noindent where $\mathcal{L}_{\text{phone}}$ denotes the Cross-Entropy (CE) loss for the auxiliary phone recognition task, and $\mathcal{L}_{mse}$ and $\mathcal{L}_{\text{conv}}$ correspond to the Mean Squared Error (MSE) and spectral convergence losses, respectively, following~\cite{mira2022svts}. In all experiments, we set $\lambda = 0.5$, consistent with our prior findings~\cite{delblanco2024comparative}.

\vspace{0.15cm}
\noindent \textbf{Voice Synthesis.} 
At inference time, the predicted Mel spectrogram,
$\hat{\mathbf{Y}}_s$, is used to synthesize the audio waveform via a pre-trained HiFTNet vocoder~\cite{li2023hiftnet}, chosen for its high-fidelity synthesis and computational efficiency.

\subsection{Multimodal Time Masking}

\noindent\textbf{Overview.} Temporal masking has previously been proposed for visual speech recognition~\cite{ma2022visual}, where contiguous time segments of the visual sequence are randomly masked during training. In this work, we extend this strategy to both sEMG and video modalities within a unified multimodal framework. Specifically, we independently apply temporal masking to each modality, mitigating over-reliance on a dominant stream and encouraging balanced cross-modal representations.

Given the mismatch in sampling rates between sEMG and video (denoted as $s_e$ and $s_v$, respectively, with $s_v \!\ll\! s_e$ in practice), we partitioned each input sequence into non-overlapping 1-second segments. This not only ensured proportional masking that is robust to varying utterance lengths, but also preserved coarse semantic alignment across modalities.

\noindent\textbf{Formulation.} Thus, for modality $m \in \{e,v\}$ with sampling rate $s_m$, each segment contains $L_m = s_m$ samples. For the $k$-th segment, we define the corresponding index set as

\begin{equation}
\mathcal{I}^{(m)}_k = \{ (k-1)L_m + 1, \dots, kL_m \}.
\end{equation}
\noindent Consistent with~\cite{ma2022visual}, the masking length is randomly sampled as $M_m \sim \mathcal{U}(0, \lfloor \rho L_m \rfloor)$, where the masking ratio, $\rho \in (0,1)$, was set to $\rho = 0.4$. When $M_m = 0$, no masking is applied. Otherwise, for $M_m > 0$, a masking start index is then sampled uniformly as $a^{(m)}_k \sim \mathcal{U}\big(1, L_m - M_m + 1\big)$, resulting in the masked index set $\mathcal{M}^{(m)}_k \subseteq \mathcal{I}^{(m)}_k$:

\begin{equation}
\mathcal{M}^{(m)}_k = \{ (k-1)L_m + a^{(m)}_k, \dots, (k-1)L_m + a^{(m)}_k + M_m - 1 \}.
\end{equation}

\noindent The consecutive frames indexed by $\mathcal{M}^{(m)}_k$ are then replaced by zero-like tensors $\mathbf{M}^{(m)}_t$ of the same shape as the original input, i.e., $\mathbf{E}_t \leftarrow \mathbf{M}^{(e)}_t, \quad \mathbf{V}_t \leftarrow \mathbf{M}^{(v)}_t, \quad \forall t \in \mathcal{M}^{(m)}_k$. Importantly, although strict temporal alignment is not enforced across modalities, both speech cues remain semantically constrained within the same 1-second segment.

\subsection{Informed Consent Statement}
All participants provided written informed consent allowing their data to be used and distributed for academic and research purposes. The informed consent procedure was approved by the Ethics Committee of the University of the Basque Country (EHU), with code M10\_2021\_269.

\section{Experimental Setup}
\label{sec:experiments}

\subsection{Data Materials}
\label{sec:data}

\begin{table*}[!t]
\centering
\caption{\textbf{Dataset Duration (hh:mm:ss) per Speaker (Gender; Age) and Split.} Values Are Reported as \textbf{Audio / EMG} Duration, Where Audio Corresponds to Audible Speech Recordings Only, Whereas EMG Includes All Electromyographic Recordings (Both Audible and Silent Sessions). 
Laryngectomized Subjects (007–009) Have No Audible Recordings.}
\label{tab:dataset}
\begin{adjustbox}{max width=1.0\textwidth}
\begin{tabular}{lcccccccccc}
\toprule & \multicolumn{6}{c}{\textbf{Laryngeal Subjects}} & & \multicolumn{3}{c}{\textbf{Alaryngeal Subjects}}\\ \cmidrule{2-7} \cmidrule{9-11}
& 001 (M; 29) & 002 (F; 29) & 003 (M; 51) & 004 (F; 46) & 005 (M; 45) & 006 (F; 61) & & 007 (F; 61) & 008 (F; 77) & 009 (M; 64)\\ \midrule

\textbf{Train} &
2:10:57 / 3:47:55 &
2:08:26 / 2:27:02 &
0:44:11 / 0:48:07 &
0:41:36 / 0:45:49 &
1:46:17 / 2:00:12 &
0:51:19 / 0:57:03 & &
\noaudio / 1:18:48 &
\noaudio / 0:26:03 &
\noaudio / 0:57:12 \\

\textbf{Dev} &
\noaudio / 0:01:42 &
\noaudio / 0:02:33 &
\noaudio / 0:00:33 &
\noaudio / 0:00:36 &
\noaudio / 0:01:56 &
\noaudio / 0:00:48 & &
\noaudio / 0:02:04 &
\noaudio / 0:00:38 &
\noaudio / 0:01:24 \\

\textbf{Test} &
\noaudio / 0:03:28 &
\noaudio / 0:05:11 &
\noaudio / 0:01:07 &
\noaudio / 0:01:11 &
\noaudio / 0:03:58 &
\noaudio / 0:01:39 & &
\noaudio / 0:04:25 &
\noaudio / 0:01:13 &
\noaudio / 0:02:52 \\
\bottomrule
\end{tabular}
\end{adjustbox}
\end{table*}

The ReSSInt dataset~\footnote{\url{https://aholab.ehu.eus/ressint/wp-content/uploads/2024/02/ReSSint\_Database_Report\_v1.pdf}}
was developed to support the creation of SSIs for Spanish speakers, specifically aiming to assist individuals who have undergone a total laryngectomy. The audio and sEMG recordings can be publicly accessed at the ELRA catalogue~\footnote{\url{https://catalog.elra.info/en-us/repository/browse/ELRA-S0498/}}, and the video is available on request. 

\begin{figure}[!t]
\centering    
\captionsetup[subfloat]{labelfont=scriptsize,textfont=scriptsize}
\subfloat[Left Side]{
\includegraphics[width=0.4\columnwidth, height=3.0cm]{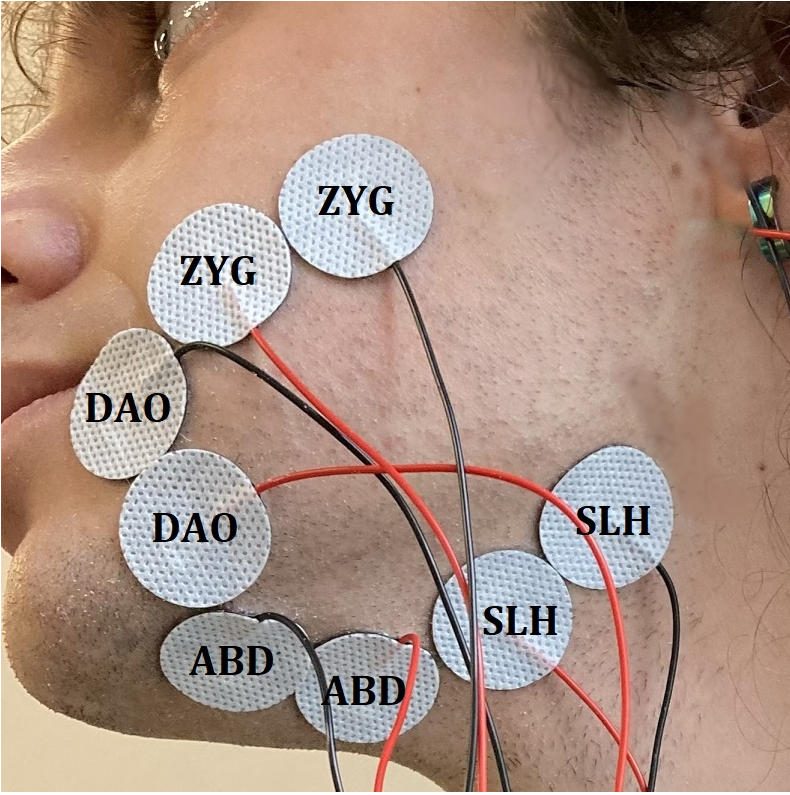}}
\subfloat[Right Side]{
\includegraphics[width=0.4\columnwidth, height=3.0cm]{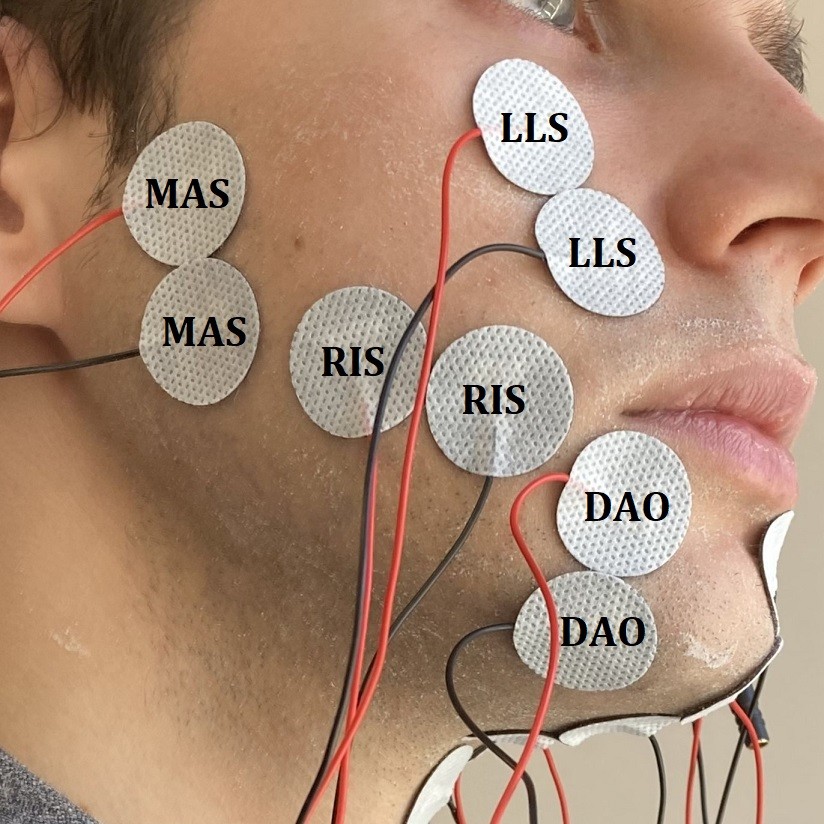}}
\caption{\textbf{Single-pair electrode setup used during the recording sessions}, comprising eight channels targeting five different facial muscles on the left (a) and right (b) sides of subject 001. Readers are referred to~\cite{salomons2025electrode} for more information about the data collection protocol.}
\label{fig:setup}
\end{figure}

\vspace{0.15cm}
\noindent\textbf{Modality \& Linguistic Content.} The participant pool therefore includes both laryngeal and post-laryngectomy individuals, who were recorded in two modalities: (i) \textbf{audible (`\texttt{a}')}
involving aloud speech; and (ii) \textbf{silent (`\texttt{s}')} where participants mouthed the text without vocalization. The dataset further defines \textbf{`\texttt{a+s}' utterances}, consisting of paired recordings where the same prompt was performed in both audible and silent modalities during the same session. In all cases, participants were asked to produce isolated words, Vowel-Consonant-Vowel (VCV) combinations, and phonemically balanced sentences, although for the purposes of this study, only the sentence-based recordings were utilized.

\vspace{0.15cm}
\noindent\textbf{Studio Setup.} In the audible modality, recordings include synchronized audio, video, and surface electromyography sEMG, while in the silent modality, recordings include video and sEMG only. To capture sEMG signals, eight bipolar sensors were placed over specific neck and facial muscles, as shown in Fig. \ref{fig:setup}, using a Quattrocento amplifier at a sampling rate of 2048 Hz. Accompanying audio was recorded at 16 kHz using a Neumann TLM103 microphone. Facial and lip movements were captured at 30 fps with a 1280$\times$720 resolution using an Intel RealSense D415 camera. All recording sessions were carried out in a controlled, quiet indoor studio environment.

\vspace{0.15cm}
\noindent\textbf{Data Split Partitioning.}
Since the model is specifically intended for use by individuals with speech impairments, the evaluation focused exclusively on silent utterances. To this end, a core corpus of 100 paired sentences (`\texttt{a+s}') was selected. From this set, 10 sentences were allocated to the development subset, 20 to the test subset, and the remaining 70 to the training set. To ensure strict text-independence, any recording of the 30 evaluation sentences (including those present only in the `\texttt{a}' modality) was excluded from the training set. All other recordings, including additional paired or single-modality utterances, were included in the training set as long as they did not violate text-independence. Due to variability in the number of sessions featuring the 100-sentence corpus across participants (three sessions for speakers 001, 002, and 005; one session for 003, 004, and 006), the resulting size of the development and test sets differs slightly between speakers. Hence, Table~\ref{tab:dataset} summarizes the data partition statistics. This includes the total audio duration, restricted to the \texttt{a} modality, and the corresponding sEMG and video signal durations, which are inherently identical. Some video recordings were occasionally missing or corrupted, resulting in slightly fewer usable samples for certain speakers.

\subsection{Comparing Modality Settings}
\label{sec:models}

To evaluate individual modality contributions, we implemented monomodal baselines that use the same encoder configuration and output heads described in section \ref{subsec:model} but bypass the fusion mechanism, enabling a direct comparison between models. Therefore, the architecture of the monomodal models consists of a convolutional frontend followed by a 6-layer E-Branchformer encoder, identical to those described in Subsection \ref{subsec:model} for each modality. Both the Mel prediction and the phone classification layers described earlier are placed directly at the encoder's output. Regarding model size, the parameter counts are 46.1M for the sEMG-based baseline, 55.6M for the video-based baseline, and 101.7M for the full multimodal model.

\subsection{Robustness to Real-World Applications}
\label{sec:robustness}

Real-world deployment of multimodal speech decoding systems often involves limited transmission bandwidth, which can degrade video and sensor signals. To simulate such conditions, we conducted experiments evaluating model performance under progressively lower bitrates, focusing specifically on the lipreading modality, which is the modality that needs a wider bandwidth and therefore is likely to degrade under suboptimal conditions. In contrast, the sEMG signal consists of only eight channels and is relatively low-dimensional; thus, it is unlikely to be significantly affected by bitrate-induced compression.

\begin{table}[!t]
\caption{\textbf{Temporal Degradation Settings} to Simulate Reduced Transmission Bitrates. Each Bitrate Corresponds to a Downsampled Lip Video Frame Rate Used During Inference.}
\label{tab:bitrates}
\footnotesize
\centering
\begin{adjustbox}{max width=0.95\columnwidth}
\begin{tabular}{@{}lcccccc@{}}
\toprule
 & \textbf{Clean} & \textbf{Mild} & \textbf{Moderate} & \textbf{Severe} & \textbf{Critical} & \textbf{Extreme} \\
\midrule
\textbf{Bitrate}     & 276  & 230  & 184  & 138  & 92  & 46 \\
\textbf{Frame Rate}  & 30   & 25   & 20   & 15   & 10  & 5 \\
\bottomrule
\end{tabular}
\end{adjustbox}
\end{table}

We first investigated whether reducing the spatial resolution of the lip ROIs impacted model performance. Specifically, we downsampled the lip ROIs from the original training resolution of $96\times96$ pixels to resolutions as low as $55\times55$ pixels, corresponding to an approximate 67\% reduction in total pixel area. Consistent with prior work in automatic lipreading~\cite{bear2014resolution,gimeno2024continuous}, we observed no significant effect of image resolution, indicating that the model is robust to spatial downsampling within the ranges considered.

Given this finding, we concentrated our analysis on temporal degradation, which is expected to have a more pronounced impact on lip-based recognition. We then reduced the effective frame rate of the lip videos to simulate progressively lower bitrates during inference. The mapping between the simulated bitrate and the corresponding frame rate is summarized in Table~\ref{tab:bitrates}. These conditions provide a controlled framework to evaluate the temporal robustness of our multimodal architectures under bandwidth-constrained scenarios.

\subsection{Implementation Details}
\label{sec:implementation}
The models were trained on a NVIDIA TITAN RTX GPU (24\,GB VRAM) and an Intel Xeon Silver 4110 CPU (8 cores, 16 threads). CUDA 13.0 was used for GPU acceleration.

\vspace{0.15cm}
\noindent\textbf{Audio Preprocessing}.
The audio modality serves both as the synthesis target and as the temporal reference used to synchronize all input streams. Audio recordings are converted into 80-bin Mel spectrograms using a 1024-sample Hanning window and a hop size of 256 samples, yielding a temporal resolution of 16 ms per frame. Afterwards, frame-level phone labels are obtained by aligning each utterance with its transcription using the Spanish Montreal Forced Aligner \cite{mcauliffe2017montreal}. Phonetic transcriptions are generated using the ahoNT\footnote{\url{https://github.com/hitz-zentroa/ahoNT}} module, resulting in a set of 29 phones plus an additional silence token, defining a total of $P = 30$ classes.

\vspace{0.15cm}
\noindent\textbf{sEMG Preprocessing}.
Consistent with the protocols described in \cite{gaddy2020digital}, we first apply a third-order Butterworth high-pass filter with a cutoff frequency of 2 Hz to suppress DC offsets and baseline drift. To mitigate power-line interference, 50 Hz notch filters with a quality factor of $Q=30$ are employed, removing the fundamental frequency and its harmonics. Furthermore, we perform data augmentation by randomly inverting the signal polarity of all eight channels with a probability of 0.5.

\vspace{0.15cm}
\noindent\textbf{Video Preprocessing}.
Following prior lipreading work~\cite{ma2022visual}, our preprocessing pipeline involves extracting $96 \times 96$ pixel grayscale patches centered on the mouth region. Face detection is performed using RetinaFace~\cite{deng2020retina}, followed by localization of 68 facial landmarks using the Face Alignment Network~\cite{bulat2017facealign}. To reduce variations in translation and scale, a similarity transformation is applied using a neutral reference frame. During training, we incorporated data augmentation through random $88 \times 88$ pixel cropping~\cite{ma2022visual}. Furthermore, since the video frame rate is lower than the temporal resolution of the target Mel spectrograms, video frames are temporally upsampled by frame repetition to match the audio frame rate.

\vspace{0.15cm}
\noindent\textbf{Training Settings}.  Following~\cite{gaddy2022voicing}, training was performed on 2-second windows rather than full utterances, improving data efficiency and training stability. Both monomodal and multimodal configurations were trained under this setting.

For the monomodal experiments, the models were first trained independently using an initial learning rate of $2.5 \cdot 10^{-4}$. The lipreading model was trained for 30 epochs, whereas the sEMG model required 50 epochs due to the higher variability and noise typically present in sEMG signals. For the multimodal experiments, we initialized the architecture using the pre-trained frontends and encoders from the monomodal models, discarding their original task-specific heads. The combined model was then trained for 5 epochs with a learning rate of $2.5 \cdot 10^{-5}$. Notably, preliminary experiments indicated that the video modality tended to dominate the multimodal learning process. To mitigate this effect, we incorporated a targeted video dropout mechanism of up to 75\%, following a strategy similar to prior audiovisual speech processing work~\cite{shi2022learning}. In all cases, optimization was performed using AdamW with a weight decay of $10^{-7}$ and a batch size of 32 samples, while a dropout rate of 0.3 was applied for regularization. The learning rate schedule included a 500-step linear warmup and was halved after 5 epochs without validation loss improvement. All hyperparameters were selected based on preliminary experimentation on the validation set.

For silent samples, we used the Mel spectrograms of the corresponding audible utterances as targets. Dynamic time warping~\cite{muller2007dynamic} is applied during training to align predictions with the reference before computing the reconstruction loss.

\vspace{0.15cm}
\noindent\textbf{Inference Procedure.} Contrary to the training procedure, for inference, the input sEMG and video sequences are not reorganized into fixed-length sequences. Thus, the context is preserved in order to keep as much information as possible.

\vspace{0.15cm}
\noindent\textbf{Evaluation Metrics.} We assess model performance from complementary perspectives. Phone Accuracy (\%Phone Acc.) is used to measure segment-level phonetic discrimination, providing insight into fine-grained alignment correctness. Word Error Rate (WER) evaluates intelligibility at the lexical level by comparing the reference text against transcriptions generated with Whisper Large-V3~\cite{radford2023robust}, thus reflecting the practical usability of the model. Finally, to quantify the structural fidelity of the reconstructed spectrogram, we compute the Structural Similarity Index (SSIM), capturing perceptually relevant differences beyond point-wise reconstruction error.
For all metrics, results are reported with 95\% confidence intervals via bootstrapping~\cite{bisani2004bootstrap}.

\section{Results \& Discussions}
\label{sec:results}

This section presents a comprehensive analysis of our experimental results\footnote{Audio samples corresponding to these experiments can be found at the following URL: \url{https://tinyurl.com/44ut6ppy
}}, where we first examine the complementarity of silent speech signals, followed by an assessment of the temporal adaptive masking strategy and its impact on multimodal learning. We then evaluate robustness under bitrate degradation settings and conduct a phone-level study, before exploring the adaptation to laryngectomized individuals.

\subsection{Multimodal Complementarity}

As discussed in the introduction, several factors motivated the design of a unified multimodal SSI capable of combining sEMG and lipreading speech cues. Among them is the limited performance achievable with sEMG alone, as evidenced by the results in Table~\ref{tab:main}, where sEMG-only models reach a WER of 83.2\% in their best configuration. In contrast, lip-only models substantially improve performance, demonstrating an intelligibility of approximately 54.5\% WER. This gap can be attributed to the lower sensitivity of lipreading signals to intra- and inter-session variability during recording, likely resulting in more stable and reliable speech representations.

Our initial hypothesis on the complementarity of both signals is then validated by the multimodal setup, which significantly outperforms single-modality models across all evaluation settings. Compared to the best lipreading-only baseline, it reduces WER by up to 14 points and improves phone-level alignment by approximately 3.5 points. This complementarity is further supported by the results in Table~\ref{tab:masking}, where the multimodal model is able to achieve better intelligibility (73.7 \% WER when the lipreading input is masked) than the sEMG-only baseline (83.2\%, Table~\ref{tab:main}). Nonetheless, the structural fidelity of the reconstructed audio appears to remain primarily driven by the lipreading modality, as reflected by its relatively small improvement with respect to the single-modality counterpart. Indeed, this observation is consistent with earlier multimodal SSI studies~\cite{freitas2014enhancing}, where visual speech cues were also found to contribute more strongly than sEMG signals in word-level recognition tasks.

\subsection{Effectiveness of the Masking Strategy}

To better understand the role of the proposed masking strategy, we analyze its impact across different modality configurations and ablation settings.

\begin{table}[!t]
\caption{\textbf{Performance Across Modalities and Ablation Settings} on the ReSSInt Test Partition for Laryngeal Subjects.}
\label{tab:main}
\footnotesize
\centering
\begin{adjustbox}{max width=0.95\columnwidth}
\begin{tabular}{@{}lccc@{}}
\toprule
\textbf{Modality / Setting} &
\textbf{\%Phone Acc.} ($\uparrow$) &
\textbf{\%WER} ($\downarrow$) &
\textbf{SSIM} ($\uparrow$) \\
\midrule

\textbf{sEMG-only} & \multicolumn{3}{l}{} \\
\addlinespace[0.3em]
Full model
 & 58.4{\tiny$\pm$0.6} & 94.1{\tiny$\pm$1.6} & 48.3{\tiny$\pm$0.5} \\
\qquad w/o Phone Loss 
 & -- & 93.6{\tiny$\pm$1.5} & 51.1{\tiny$\pm$0.4} \\
\qquad w/o Random Masking
 & 63.6{\tiny$\pm$0.6} & 83.2{\tiny$\pm$1.8} & 52.0{\tiny$\pm$0.5} \\

\addlinespace[0.6em]
\textbf{Lip-only} & \multicolumn{3}{l}{} \\
\addlinespace[0.3em]
Full model
 & 71.4{\tiny$\pm$0.6} & 57.2{\tiny$\pm$2.0} & 54.8{\tiny$\pm$0.4} \\
\qquad w/o Phone Loss 
 & -- & 62.0{\tiny$\pm$2.1} & 56.1{\tiny$\pm$0.4} \\
\qquad w/o Random Masking
 & 72.8{\tiny$\pm$0.6} & 54.5{\tiny$\pm$2.1} & 56.8{\tiny$\pm$0.4} \\

\addlinespace[0.6em]
\rowcolor{gray!25}
\textbf{sEMG + Lips (Multimodal)} & & & \\
\rowcolor{gray!12}
Full Model
 & 76.3{\tiny$\pm$0.7} & 40.5{\tiny$\pm$2.8} & 56.6{\tiny$\pm$0.6} \\
\rowcolor{gray!12}
\qquad w/o Phone Loss 
 & -- & 58.8{\tiny$\pm$2.2} & 55.5{\tiny$\pm$0.4} \\
\rowcolor{gray!12}
\qquad w/o Random Masking
 & 76.2{\tiny$\pm$0.5} & 44.0{\tiny$\pm$2.1} & 57.2{\tiny$\pm$0.4} \\

\bottomrule
\end{tabular}
\end{adjustbox}
\end{table}

\vspace{0.15cm}
\noindent\textbf{Multimodal-Dependent Effect.} Importantly, the proposed temporal adaptive masking strategy is only effective in the multimodal setting. While it improves intelligibility when both modalities are present (although improvement seems not statistically significant), its incorporation degrades performance in unimodal configurations, as reflected by the ablation results in Table~\ref{tab:main}. For lipreading, this results in a not significant degradation of approximately 2.7 WER points, whereas the effect is highly significant for sEMG, with a deterioration of around 10 WER points when masking is applied. However, masking has been shown to be effective in unimodal lipreading-to-text models~\cite{ma2022visual}, which suggests that its benefits do not directly transfer to speech reconstruction settings such as our case study, where the model must generate continuous acoustic structures rather than discrete linguistic units. A possible explanation is that, in unimodal settings, masking removes essential information that cannot be easily recovered in this more challenging generation task without complementary cues.

Overall, these findings support the hypothesis that such masking promotes more robust cross-modal learning, encouraging the model to effectively leverage complementary information across modalities, in line with prior work in audio-visual speech processing~\cite{shi2022learning}. We further evaluated alternative masking strategies, including half-face and channel-wise sEMG masking aligned with corresponding and opposite facial regions; however, these variants did not lead to consistent improvements or stable trends across settings.

\vspace{0.15cm}
\noindent\textbf{Limitations of Corruption-Specific Training.} Despite the observations so far, it remains unclear whether the observed robustness stemmed from the masking strategy itself or merely from exposure to corrupted inputs during training. To disentangle these factors, we trained an additional model using frame-rate augmentation. As shown in Table~\ref{tab:masking}, although this approach is competitive with the masking-based model when one modality is missing, it yields significantly weaker performance under clean (unmasked) conditions. This deterioration, together with the observation that masking the sEMG modality at test time improves performance, suggests that corruption-specific augmentation does not effectively promote multimodal fusion. Rather than learning complementary cross-modal representations, this type of augmentation appears to bias the model toward learning to denoise or compensate for the specific degradation encountered during training---a phenomenon also found in the domain of audio-visual speech recognition when introducing noise in the more informative audio signal~\cite{gimeno2025tailored}. In contrast, random masking demonstrates a more balanced and stable multimodal integration, as evidenced by our previous experimental analyses. These findings thus confirm our hypothesis that a general masking strategy provides a robust and generalizable approach to handling unexpected or diverse signal corruptions, without requiring the design of task-specific augmentation schemes.

\begin{table}[!t]
\caption{\textbf{Robustness to Missing Modalities} of the sEMG+Lips Model Trained with Different Data Augmentation Strategies.}
\label{tab:masking}
\footnotesize
\centering
\begin{adjustbox}{max width=\columnwidth}
\begin{tabular}{@{}lccc@{}}
\toprule
& \multicolumn{3}{c}{\textbf{Test-Time Inference} (\texttt{\%WER ($\downarrow$) / \pacc{\%Phone Acc. ($\uparrow$)})}}\\ \cmidrule{2-4}
\textbf{Training Setting}
 & \textbf{No Mask}
 & \textbf{Masked sEMG}
 & \textbf{Masked Lips} \\
\midrule

w/o Random Masking 
 & 44.0{\tiny$\pm$2.1} / \pacc{76.2{\tiny$\pm$0.5}}
 & 46.9{\tiny$\pm$2.0} / \pacc{69.1{\tiny$\pm$0.6}}
 & 92.5{\tiny$\pm$1.6} / \pacc{62.5{\tiny$\pm$0.6}}\\

{w/ Frame-Rate Aug.}
 & 57.3{\tiny$\pm$2.2} / \pacc{73.9{\tiny$\pm$0.5}}
 & 50.9{\tiny$\pm$2.1} / \pacc{49.6{\tiny$\pm$0.7}}
 & 77.5{\tiny$\pm$1.9} / \pacc{64.4{\tiny$\pm$0.6}}\\ 
 
\rowcolor{gray!12}
w/ Random Masking
 & 40.5{\tiny$\pm$2.8} / \pacc{76.3{\tiny$\pm$0.7}}
 & 48.2{\tiny$\pm$2.0} / \pacc{74.5{\tiny$\pm$0.6}}
 & 73.7{\tiny$\pm$4.4} / \pacc{65.1{\tiny$\pm$1.2}}\\

\bottomrule
\end{tabular}
\end{adjustbox}
\end{table}

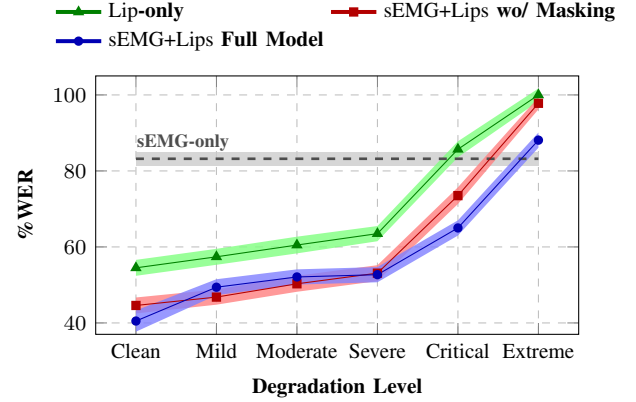
\begin{figure}[!t]
\centering
\begin{tikzpicture}
    \begin{axis}[
        width=0.9\columnwidth,
        height=5cm,
        xlabel={\textbf{Degradation Level}},
        ylabel={\textbf{\%WER}},
        xtick={1,2,3,4,5,6},
        xticklabels={Clean, Mild, Moderate, Severe, Critical, Extreme},
        ymin=37, ymax=105,
        ymajorgrids=true,
        grid=both,
        grid style=dashed,
        legend style={
            at={(0.5,1.05)},
            anchor=south,
            legend columns=2,
            legend cell align=left,
            font=\footnotesize,
            draw=none
        },
        line width=1.5pt,
        every axis label/.append style={font=\footnotesize},
        tick label style={font=\footnotesize}
    ]

    \addplot[name path=emg_upper, draw=none, forget plot] coordinates {
        (1,83.2+1.8) (6,83.2+1.8)
    };
    \addplot[name path=emg_lower, draw=none, forget plot] coordinates {
        (1,83.2-1.8) (6,83.2-1.8)
    };
    \addplot [black!40, fill=black!20, opacity=0.8, forget plot] fill between[
        of=emg_upper and emg_lower
    ];

    \addplot[name path=lip_upper, draw=none, forget plot] coordinates {
        (1, 54.5+2.1) (2, 57.4+2.1) (3, 60.5+2.2) (4, 63.5+2.0) (5, 85.7+2.1) (6, 100.0+1.7)
    };
    \addplot[name path=lip_lower, draw=none, forget plot] coordinates {
        (1, 54.5-2.1) (2, 57.4-2.1) (3, 60.5-2.2) (4, 63.5-2.0) (5, 85.7-2.1) (6, 100.0-1.7)
        
    };
    \addplot [green!40, fill=green!50, opacity=0.8, forget plot] fill between[
        of=lip_upper and lip_lower
    ];

    \addplot[name path=full_upper, draw=none, forget plot] coordinates {
        (1, 44.6+2.1) (2, 46.8+2.0) (3, 50.3+2.1) (4, 53.1+2.1) (5, 73.5+2.2) (6, 97.8+1.6)
    };
    \addplot[name path=full_lower, draw=none, forget plot] coordinates {
        (1, 44.6-2.1) (2, 46.8-2.0) (3, 50.3-2.1) (4, 53.1-2.1) (5, 73.5-2.2) (6, 97.8-1.6)
    };
    \addplot [red!40, fill=red!50, opacity=0.8, forget plot] fill between[
        of=full_upper and full_lower
    ];

    \addplot[name path=mask_upper, draw=none, forget plot] coordinates {
        (1, 40.5+2.8) (2, 49.4+2.1) (3, 52.1+2.0) (4, 52.7+2.0) (5, 65.0+2.0) (6, 88.1+2.0)
    };
    \addplot[name path=mask_lower, draw=none, forget plot] coordinates {
        (1, 40.5-2.8) (2, 49.4-2.1) (3, 52.1-2.0) (4, 52.7-2.0) (5, 65.0-2.0) (6, 88.1-2.0)
    };
    \addplot [blue!40, fill=blue!50, opacity=0.8, forget plot] fill between[
        of=mask_upper and mask_lower
    ];

    \addplot[name path=emg, color=black!70, dashed, line width=1.pt, forget plot]
    coordinates {
        (1,83.2) (6,83.2)
    };
    \node[
        anchor=west,
        font=\scriptsize,
        text=black!70
    ] at (axis cs:0.9,87.5) {\textbf{sEMG-only}};

    \addplot[name path=lip, color=green!50!black, mark=triangle*, mark size=1.pt]
    coordinates {
        (1, 54.5) (2, 57.4) (3, 60.5) (4, 63.5) (5, 85.7) (6, 100.0)
    };
    \addlegendentry{Lip\textbf{-only}}

    \addplot[name path=nomask, color=red!70!black, mark=square*, mark size=1.pt] coordinates {
        (1, 44.6) (2, 46.8) (3, 50.3) (4, 53.1) (5, 73.5) (6, 97.8)
    };
    \addlegendentry{sEMG+Lips \textbf{wo/ Masking}}
    
    \addplot[name path=mask, color=blue!70!black, mark=*, mark size=1.pt] coordinates {
        (1, 40.5) (2, 49.4) (3, 52.1) (4, 52.7) (5, 65.0) (6, 88.1)
    };
    \addlegendentry{sEMG+Lips \textbf{Full Model}}

    \end{axis}
\end{tikzpicture}
\caption{\textbf{Robustness across increasing temporal degradation levels.} For our full multimodal model and its ablation variants, shaded bands denote 95\% confidence intervals. Single-modality baselines are included for reference. A formal definition of the different degradation levels can be found in Table~\ref{tab:bitrates}.}
\label{fig:degradation}
\end{figure}

\subsection{Robustness to Temporal Degradation Settings}

To simulate real-world deployment under limited transmission bandwidth, we evaluate robustness to progressive frame-rate reduction of the lipreading modality. Beyond the previously discussed multimodal gains, Fig.~\ref{fig:degradation} reveals a remarkable robustness of lipreading-based approaches to temporal degradation. From \textit{clean} to \textit{severe} conditions, although exhibiting a gradual performance decline, both multimodal variants nearly follow identical trends, indicating stable and predictable behavior under moderate frame-rate reduction. However, once the degradation surpasses the \textit{critical} threshold (15 fps), significant differences emerge. In particular, removing the masking strategy results in a pronounced performance drop, highlighting the importance of masking-based training to promote effective multimodal integration and maintain competitive performance relative to the sEMG-only counterpart.

\subsection{Phone-Level Error Analysis}

\begin{figure}[!t]
    \centering

    \makeatletter\def\@captype{figure}\makeatother
    
    \includegraphics[width=1.0\columnwidth]{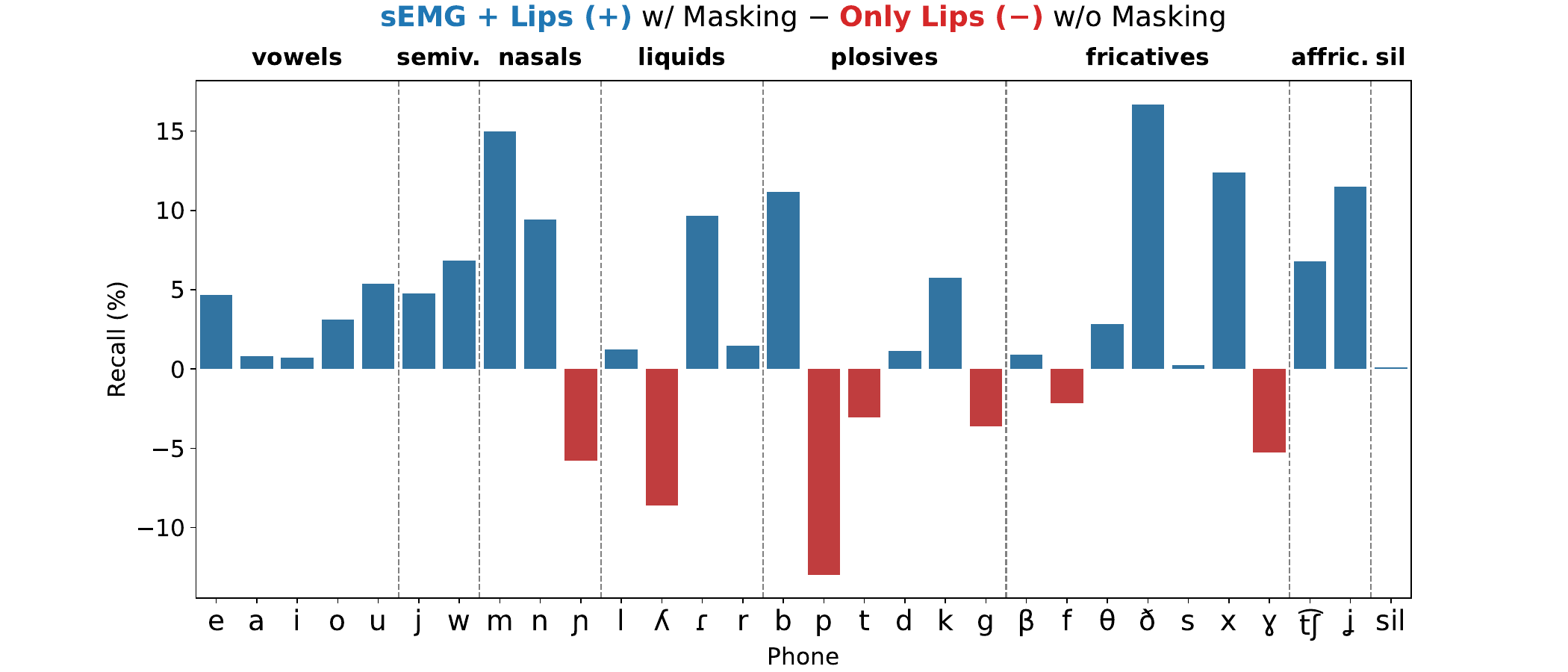}
    \caption{\textbf{Phone-level differential analysis} between the multimodal model with full masking and the monomodal video-based model trained without masking. Positive blue bars indicate that the recall for that phone is higher in the multimodal model with respect to the one based solely on lip-reading.}
    \label{fig:phone-analysis}
    
    \vspace{1.5em}
    \makeatletter\def\@captype{table}\makeatother
    
    \caption{\textbf{Comparison of the Most Frequent Phone Confusion Pairs} Between the Lipreading and Multimodal Models}
    \label{tab:comparative-confusion}
    \small 
    \begin{adjustbox}{max width=0.8\columnwidth}
    \begin{tabular}{lccc}
        \toprule
        \multirow{2}{*}[-4pt]{\makecell{\textbf{Confusion Type}\\(True $\rightarrow$ Pred.)}} & \multicolumn{2}{c}{\textbf{Phone Confusion (\%)}} & \multirow{2}{*}[-4pt]{\textbf{$\Delta$}} \\ 
        \cmidrule{2-3}
        & \textbf{Lips-Only} & \textbf{sEMG + Lips} & \\
        \midrule
        
        \addlinespace[0.6em]
        \rowcolor{gray!12}
        \textit{\textbf{Nasal confusions}} & & & \\
        \addlinespace[0.3em]
        b $\rightarrow$ m & 12.1 & 19.5 & +7.4 \\
        p $\rightarrow$ m & 12.5 & 19.5 & +7.0 \\
        g $\rightarrow$ n & 10.9 & 12.2 & +1.3 \\
        
        \addlinespace[0.6em]
        \rowcolor{gray!12}
        \textit{\textbf{Voicing confusions}} & & & \\
        \addlinespace[0.3em]
        b $\rightarrow$ p & 29.6 & 16.3 & -13.3 \\
        d $\rightarrow$ t & 14.2 & 13.5 & -0.7 \\
        g $\rightarrow$ k & 15.8 & 19.5 & +3.6 \\
        \bottomrule
    \end{tabular}
    \end{adjustbox}

    \vspace{-0.2cm}

\end{figure}

We further analyze phone-level performance by comparing the best-performing configurations of the multimodal model and the lipreading-only baseline on the ReSSInt dataset. As Fig.~\ref{fig:phone-analysis} depicts, we computed per-phone recall differences and group phones by articulatory class to investigate how sEMG complemented visual speech cues.  All phones shown in the referenced figure are represented using the International Phonetic Alphabet (IPA). 

Overall, these results indicate that multimodal fusion leads to consistent improvements in phone recognition across most articulatory classes. The most pronounced gains are observed for vowels and affricates, suggesting that sEMG provides complementary information. Improvements are also visible in several consonantal categories; however, a more mixed behavior emerges for plosives and nasals, where gains are less uniform across individual phones. Notably, silence remains largely unaffected, which is expected given its role as a boundary token occurring primarily at the beginning and at the end of utterances rather than as an active speech segment.

The confusion analysis in Table~\ref{tab:comparative-confusion} complements this analysis by reporting the most frequent phone confusion pairs shared between both models, providing an overview of their most persistent error patterns. The results highlight the challenges of working with signals that are inherently ambiguous or only partially informative for capturing the full range of speech production cues. A representative example of this is the set of confusions between phones such as $b$, $p$, and $m$, which correspond to visually similar articulatory units, i.e., visemes~\cite{fernandez2017automatic}. Only in the cases involving voicing-related contrasts, such as $b \rightarrow p$ and $d \rightarrow t$, we observe improvements as a result of multimodal fusion, suggesting that sEMG provides complementary information about internal speech movements that cannot be captured by video, even in silent mode. However, the confusion between $g \rightarrow k$ is larger with the multimodal fusion. This may be due to the articulation point of those phones being the soft palate, whose movement is hardly registered by the surface sensors. Consequently, the sEMG signal fails to provide discriminative features for these phones, acting instead as a source of noise that interferes with the visual cues. Overall, these results suggest that while incorporating sEMG is beneficial for specific articulatory contrasts, its contribution is not uniform across all phonetic classes.

\subsection{Adaptation to Laryngectomized Subjects}

Even when training on silent speech samples, SSI systems are typically developed using laryngeal speakers~\cite{gaddy2022voicing,petridis2018silent}, for whom paired audible utterances from the same speaker are available to provide supervision and temporal alignment targets during training. In contrast, adaptation to laryngectomized individuals poses fundamental challenges, as no corresponding natural voice recordings exist for the target speaker. As a result, training must rely on surrogate reference signals or cross-speaker alignment strategies, where differences in articulation manner or speech pace may hinder accurate adaptation.

This inherent mismatch is likely responsible for the performance degradation observed in Table~\ref{tab:alaryngeal}, where the best-performing model for each modality setting was fine-tuned for up to 20 epochs using a learning rate of $2.5 \times 10^{-5}$. For this adaptation, we conduct experiments using laryngeal speakers as surrogate reference voices, specifically selecting subjects 001, 002, and 005, which contain the largest amount of available data (see Table~\ref{tab:dataset}). Among these, the reference speaker yielding the best performance in terms of WER for each laryngectomized target was selected as the final alignment reference, which was achieved in all cases using subject 002 as target voice. Interestingly, the multimodal configuration does not appear to significantly outperform the lipreading-only model in this case study. This may be attributed to the strong degradation observed in sEMG-only performance, which can negatively affect the fusion process by introducing unreliable or noisy representations into the multimodal model.

\begin{table}[!t]
\caption{\textbf{Performance Across Modalities} for the Best-Performing Models Adapted to the ReSSInt Laryngectomized Subjects.}
\label{tab:alaryngeal}
\footnotesize
\centering
\begin{adjustbox}{max width=0.9\columnwidth}
\begin{tabular}{@{}lccccc@{}}
\toprule
\textbf{Modality} &  &
\textbf{\%Phone Acc.} ($\uparrow$) &
\textbf{\%WER} ($\downarrow$) &
\textbf{SSIM} ($\uparrow$) \\
\midrule

\textbf{sEMG-only} & & 44.3{\tiny$\pm$1.5} & 90.9{\tiny$\pm$2.7} & 39.5{\tiny$\pm$0.5} \\

\addlinespace[0.6em]
\textbf{Lip-only} & & 62.2{\tiny$\pm$1.1} & 75.1{\tiny$\pm$3.7} & 46.7{\tiny$\pm$0.6} \\

\addlinespace[0.6em]
\textbf{sEMG + Lips} & & 63.6{\tiny$\pm$1.2} & 68.2{\tiny$\pm$3.3} & 46.7{\tiny$\pm$0.5} \\

\bottomrule
\end{tabular}
\end{adjustbox}
\end{table}

\begin{figure}[!t]
\centering
\begin{tikzpicture}
\begin{axis}[
    width=0.9\columnwidth,
    height=5cm,
    ylabel={\textbf{\%WER}},
    xlabel={\textbf{Subject ID}},
    symbolic x coords={007, 008, 009},
    xtick=data,
    ymin=40, ymax=116,
    ymajorgrids=true,
    grid=both,
    grid style=dashed,
    legend style={
        at={(0.5,1.05)},
        anchor=south,
        legend columns=3,
        font=\footnotesize,
        draw=none
    },
    ybar=6pt,
    bar width=6pt,
    enlarge x limits=0.2,
    every axis label/.append style={font=\footnotesize},
    tick label style={font=\footnotesize},
]

\addplot[
    ybar,
    fill=black!50,
    draw=black!70,
    error bars/.cd,
        y dir=both,
        y explicit
] coordinates {
    (007,98.8) +- (0,2.5)
    (008,108.7) +- (0,4.7)
    (009,74.1) +- (0,4.6)
};
\addlegendentry{sEMG\textbf{-only}}

\addplot[
    ybar,
    fill=green!50,
    draw=green!50!black,
    postaction={
        pattern=north west lines,
        pattern color=black
    },
    error bars/.cd,
        y dir=both,
        y explicit
] coordinates {
    (007,93.4) +- (0,5.2)
    (008,70.1) +- (0,7.8)
    (009,59.3) +- (0,5.2)
};
\addlegendentry{Lip\textbf{-only}}

\addplot[
    ybar,
    fill=blue!50,
    draw=blue!40!black,
    postaction={
        pattern=north east lines,
        pattern color=black,
    },
    error bars/.cd,
        y dir=both,
        y explicit
] coordinates {
    (007,84.2) +- (0,3.7)
    (008,63.1) +- (0,7.1)
    (009,54.8) +- (0,5.4)
};
\addlegendentry{sEMG+Lips \textbf{Full}}

\end{axis}
\end{tikzpicture}
\caption{\textbf{Per-subject performance across modalities} on the ReSSInt test partition for laryngectomized subjects, showing inter-speaker variability.}
\label{fig:persubject}
\end{figure}
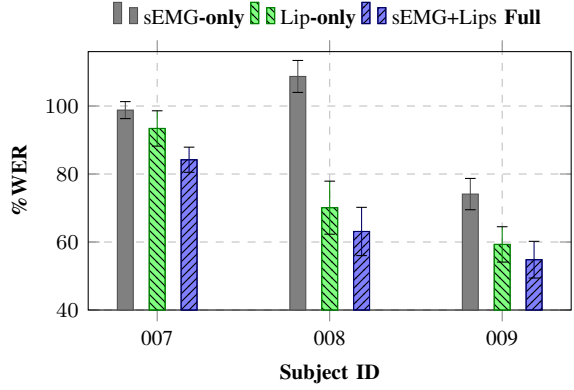

\begin{figure}[!t]
    \centering
    \includegraphics[trim=1cm 0cm 1cm 0cm, clip, width=\linewidth]{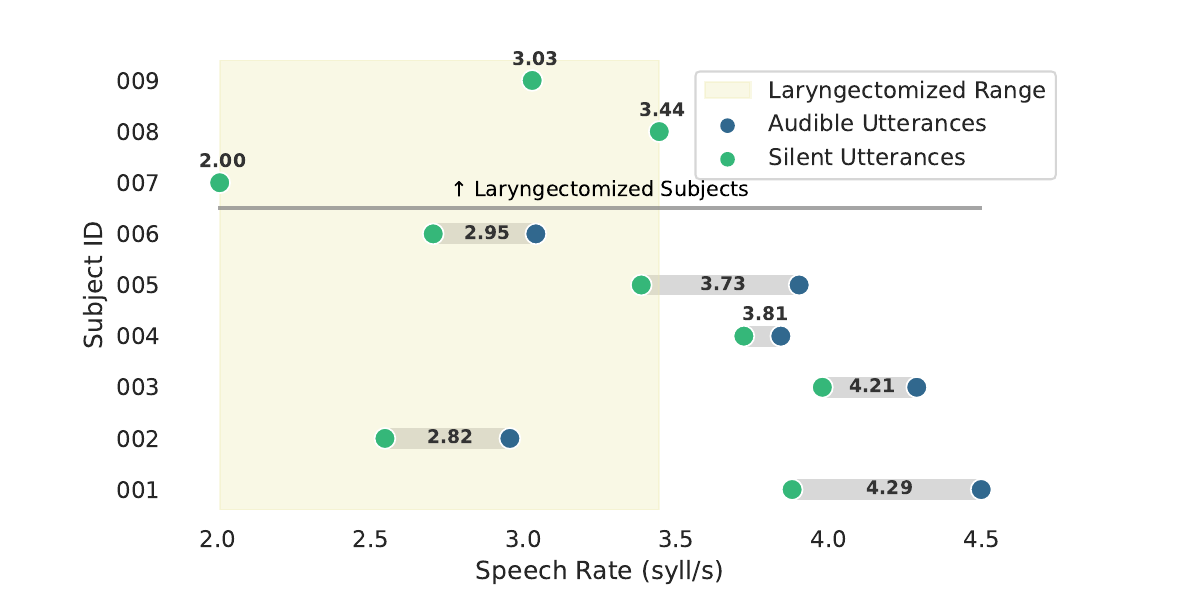}
    \caption{\textbf{Comparison of average speech rates across speakers} indicating the average speech rate in syllables per second (syll/s). The shaded area highlights the speech-rate range observed for laryngectomized subjects.}
    \label{fig:speech-rate}
\end{figure}

Another important aspect is that post-laryngectomy speech production is highly heterogeneous. Depending on the rehabilitation strategy, individuals may rely on electrolarynx devices or esophageal speech (all alaryngeal subjects in our dataset use esophageal speech), both of which involve learning substantially altered articulatory dynamics that can deviate from pre-operative speech patterns. Consequently, the mismatch between silent articulatory signals and surrogate laryngeal target speech can become substantially more pronounced, making inter-speaker variability a dominant factor, as reflected in Fig.~\ref{fig:persubject}.

To further investigate this variability, we analyzed the speech rate (measured as syllables per second) across all subjects, distinguishing between audible and silent utterances (see Fig. \ref{fig:speech-rate}).

Our observations indicate that the speech rate of subject 002 is the most similar to those of the laryngectomized subjects, which could explain why this speaker consistently provided the most effective adaptation targets. Conversely, the exceptionally low speech rate of laryngectomized subject 007 may explain the poorer performance observed for this individual during adaptation. Regarding the performance gap between subjects 008 and 009, the higher error rates for the former may be attributed to the more limited amount of data available for subject 008. Together, this speaker variability underscores the complexity of the task.

\section{Conclusion \& Future Works}

This work addresses a current research gap in multimodal silent speech interfaces by designing a speech synthesis system that combines sEMG and lipreading cues. Our experiments confirm the complementarity of both silent signals and demonstrate that temporal adaptive masking plays a key role in their effective integration. The robustness and generalization capabilities of this strategy are supported by additional experiments comparing its effect with frame-rate distortions, which would avoid the design of task-specific augmentation schemes, and bitrate degradations simulating real-world case scenarios. Additional phone-level analysis provides further insights into which articulatory classes benefit most from multimodal fusion, demonstrating that vowels and affricates exhibit the greatest improvements.

Nevertheless, our results also show that adaptation to laryngectomized individuals remains challenging, likely due to the substantial inter-speaker variability associated with post-surgical speech production, rehabilitation strategies, and compensatory articulatory behaviors. In this context, a promising direction for future work is the exploration of the recently introduced soft speech units~\cite{scheck2023multi}, which learn to map a single user’s sEMG data to synthesize target audio from multiple target voices. By encouraging speaker-independent representations, such units may help reduce the impact of this variability and provide a more flexible latent space for speech reconstruction within our multi-speaker pretraining framework prior to adaptation. Moreover, extending this paradigm with synthetic speech could further increase speaker diversity during pretraining or even enable the generation of target speech better matched to the silent articulatory patterns of the laryngectomized individual~\cite{scheck2025diffmvets}. In this sense, given our findings suggesting that differences in speech pace may influence adaptation performance, the investigation of data augmentation strategies that explicitly vary speech rate in the audio, visual, and sEMG modalities could prove beneficial.

From a multimodal learning perspective, future work will also explore knowledge distillation strategies across modalities, with the goal of improving the weaker performance of the EMG-only branch compared to lipreading. In particular, prior work~\cite{afouras2018deep,ma2021lira} has shown that visual speech models can benefit from audio-guided or cross-modal distillation, suggesting that similar strategies may help transfer complementary information from stronger modalities to EMG-based representations. Another potential line involves enriching these cross-modal articulatory representations through diffusion modeling, as successfully demonstrated in the context of vocal tract Magnetic Resonance Imaging (MRI) synthesis~\cite{perez2026cross}.

\section*{Acknowledgments}
The work of del Blanco, Navas, and Hernáez is part of the project PID2022-141378OB-C21, funded by MCIN/AEI/ 10.13039/501100011033/FEDER,UE, and supported by FPI grant PREP2022-000130. The work of Gimeno-Gómez and Martínez-Hinarejos was partially supported by the GVA (Grant CIACIF/2021/295) and the PROMETEO 2024 program (project LightVED, CIPROM/2023/17), forming part also of the R\&D\&I project ANNOTATE-MULTI2 (PID2024-156022OB-C32), funded by MICIU/AEI and FEDER/EU, and of the Iberian Digital Media Observatory (IBERIFIER Plus), co-funded by the EC under Call DIGITAL2023-DEPLOY-04 (Grant 101158511). Finally, we would also like to thank everyone who took part in this study.

\bibliographystyle{IEEEtran}
\bibliography{main}

\vspace{-22pt}
\begin{IEEEbiography}[{\includegraphics[width=1in,height=1.25in,clip,keepaspectratio]{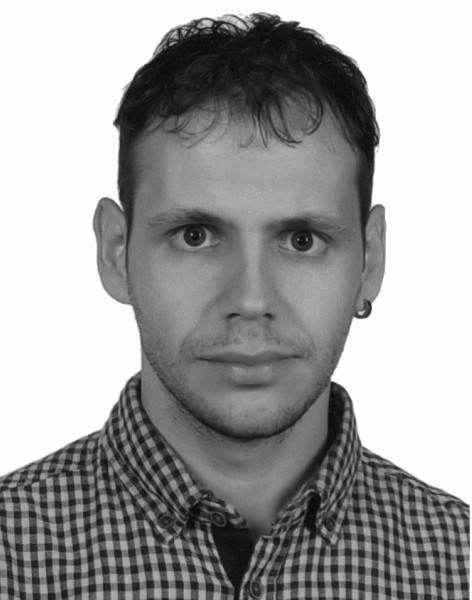}}]{Eder del Blanco} received the B.Sc. degree in Telecommunication Technology (2015) and the M.Sc. degree in Telecommunication Engineering (2020) from the University of the Basque Country (UPV/EHU), Bilbao, Spain. He is currently pursuing the Ph.D. degree in Language Analysis and Processing at UPV/EHU. He has been a visiting researcher at the Dalle Molle Institute for Artificial Intelligence (IDSIA) of the USI-SUPSI, Lugano, Switzerland, under the supervision of Prof. Michael Wand. His current research focuses on multimodal silent speech interfaces.
\end{IEEEbiography}

\vspace{-22pt}
\begin{IEEEbiography}[{\includegraphics[width=1in,height=1.25in,clip,keepaspectratio]{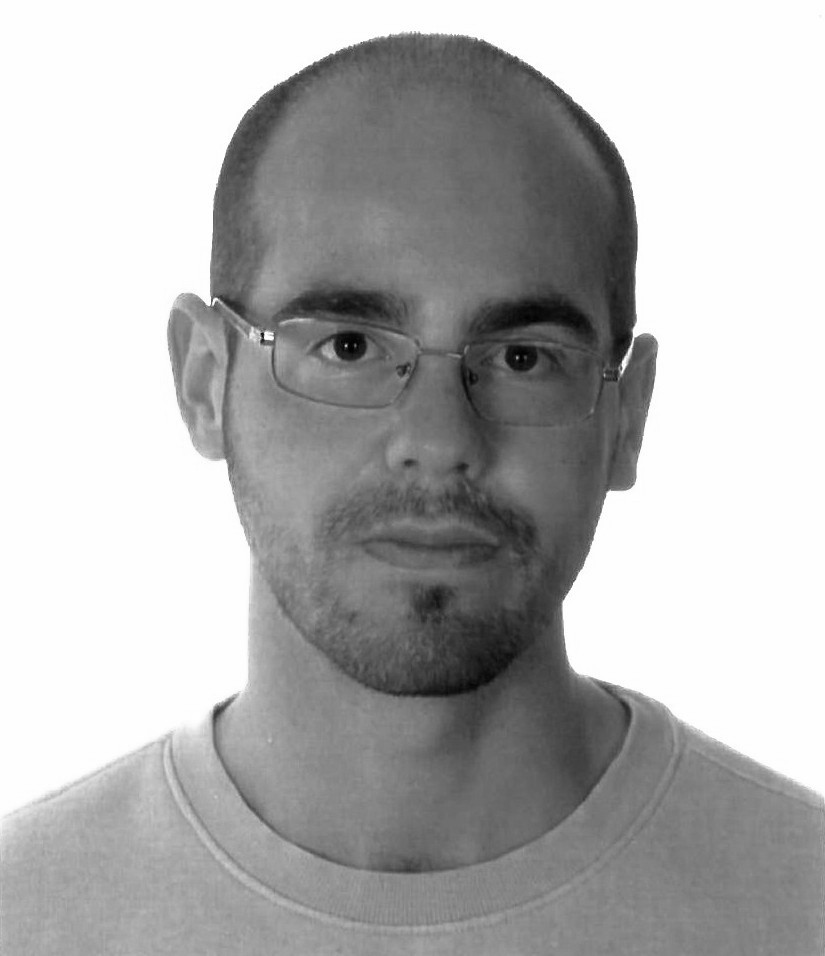}}]{David Gimeno-Gómez} received the B.Sc. degree in computer science and the M.Sc. degree in artificial intelligence, pattern recognition, and digital image from the Universitat Politècnica de Valéncia (UPV), Spain, in 2019 and 2020, respectively. He completed his Ph.D in computer science at UPV in 2025, with International Cum Laude distinction. He has been a visiting research at HLT group of INESC-ID in Lisbon (Portugal) under the supervision of Prof. Alberto Abad. His research topics currently focus on multimodal speech technologies, with particular interest in silent speech interfaces and pathological speech analysis.
\end{IEEEbiography}

\vspace{-22pt}
\begin{IEEEbiography}[{\includegraphics[width=1in,height=1.25in,clip,keepaspectratio]{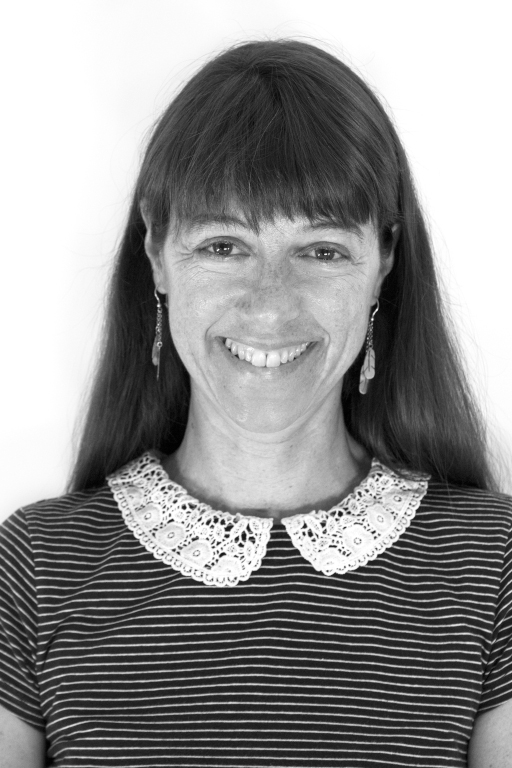}}]{Eva Navas} was born in San Sebastián, Spain, in 1971. She received the Telecommunication Engineering and Ph.D. degrees from the Department of Electronics and Tele-communications, University of the Basque Country, Bilbao, Spain, in 1996 and 2003, respectively.,She is currently a Researcher with the Aholab Signal Processing Laboratory, Department of Communications Engineering, University of the Basque Country, where she teaches with the Faculty of Engineering. She works mainly in diarization and speech synthesis.
\end{IEEEbiography}

\vspace{-22pt}
\begin{IEEEbiography}[{\includegraphics[width=1in,height=1.25in,clip,keepaspectratio]{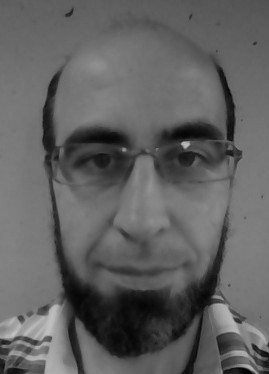}}]{Carlos-D. Martínez-Hinarejos} received the B.Sc. degree in computer science, the Ph.D. degree in pattern recognition and artificial intelligence, and the B.Sc. degree in biotechnology from the Universitat Politècnica de València (UPV), Valencia, Spain, in 1998, 2003, and 2012, respectively. He joined the UPV staff in the Computation and Computer Systems Department, UPV, in 2000. He pertains to the Pattern Recognition and Human Language Technology Research Center, where he develops his research on the topics of speech recognition, dialogue systems, multimodal systems, and text classification. He has participated in many European and Spanish projects, and is an active member of the Spanish Network for Speech Technologies.
\end{IEEEbiography}

\vspace{-22pt}
\begin{IEEEbiography}[{\includegraphics[width=1in,height=1.25in,clip,keepaspectratio]{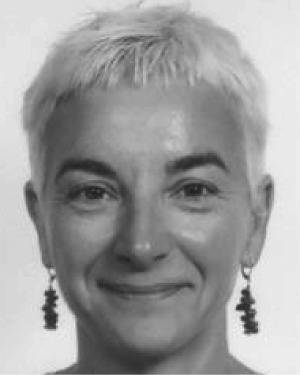}}]{Inma Hernáez} received the Telecommunication Engineering degree from the Universitat Politecnica de Catalunya, Barcelona, Spain, in 1987, and the Ph.D. degree in Telecommunication Engineering from the University of the Basque Country, Bilbao, Spain, in 1995. She is currently a Full Professor with the Department of Communications Engineering, Faculty of Engineering, University of the Basque Country, where she is involved in the area of signal theory and communications, and the Founding Member and Director of the Aholab Signal Processing Laboratory. Her research interests include signal processing and all aspects related to speech processing. She is also interested in the development of speech resources and technologies for the Basque language.
\end{IEEEbiography}

\vfill

\end{document}